\documentclass[a4paper]{easychair}
\usepackage{amsmath,amssymb,amsfonts}
\usepackage{algorithmic}
\usepackage{graphicx}
\usepackage{textcomp}
\usepackage{csquotes}
\usepackage{lineno}
\usepackage{listings}
\usepackage{xcolor}
\usepackage{framed}
\usepackage{empheq}
\usepackage{mwe}
\usepackage{caption}
\usepackage{subcaption}
\usepackage{verbatim}
\usepackage{pgfplots}
\usepgfplotslibrary{patchplots}
\usepackage{algorithm}
\usepackage{algorithmic}
\usepackage{enumitem}

\definecolor{lightergray}{gray}{0.95}

\colorlet{shadecolor}{lightergray}

\usepackage{todonotes}
\newboolean{showcomments}
\setboolean{showcomments}{true} 
\ifthenelse{\boolean{showcomments}}
{\newcommand{\nb}[2]{
  \fcolorbox{black}{yellow}{\bfseries\sffamily\scriptsize#1}
  {\sf\small$\blacktriangleright$\textit{#2}$\blacktriangleleft$}
 }
 
}
{\newcommand{\nb}[2]{}
 
}
\newcommand\christos[1]{\nb{Christos}{#1}}

\usepackage{wrapfig}
\usepackage{mathtools}

\usepackage{fancyhdr}
\begin{document}


\title{
Early Validation of Cyber-Physical Space Systems via Multi-Concerns Integration
}


\author{
    Nianyu Li\inst{1}
\and
    Christos Tsigkanos\inst{2}
\and
    Zhi Jin\inst{1}
\and
   Zhenjiang Hu\inst{1}
\and 
    Carlo Ghezzi\inst{3}
}
\institute{
  Key Laboratory of High Confidence Software Technologies (MoE), Peking University, China\\
\and
   Distributed Systems Group, TU Wien, Austria\\
\and
   Dipartimento di Elettronica Informazione e Bioingegneria, Politecnico di Milano, Italy\\
 }

\maketitle

\begin{abstract}
Cyber-physical space systems are engineered systems operating within physical space with design requirements that depend on space, e.g., regarding location or movement behavior. They  are built from and depend upon the seamless integration of computation and physical components. Typical examples include systems where software-driven agents such as mobile robots explore space and perform actions to complete particular missions. Design of such a system often depends on multiple concerns expressed by different stakeholders, capturing different aspects of the system. We propose a model-driven approach supporting  (a) separation of concerns during design, (b) systematic and semi-automatic integration of separately modeled concerns, and finally (c) early validation via statistical model checking. We evaluate our approach over two different case studies of cyber-physical space systems.


\end{abstract}



\maketitle


\section{Introduction}
Cyber-Physical Space Systems (CPSS) are an important class of cyber-physical systems (CPS), the term refers to the tight integration of and coordination between computational and physical resources.  
A CPSS is a CPS deployed in a physical space, which exhibits functionalities that depend on the structure of the space and on physical locations inherent in it. In this paper, we explicitly focus on CPSS inhabited by human and autonomous agents, which need to accomplish certain  missions in space. For example, robots move goods between different locations in an office space, or UAVs try to rescue victims in a disaster recovery scenario.



Modeling and validation have been acknowledged as critical activities in systematic system design~\cite{MDPmodelchecking}. 
Model construction is especially hard and challenging in the case of CPSS where many diverse aspects are intertwined~\cite{fse17}. 
By exploring a variety of cases, we can see that some of the typical aspects include the spatial domain in which the system will be deployed, the allowed movements of the system entities in space, their interaction and cooperation strategies as well as the mission that the system needs to accomplish~\cite{DBLP:journals/pieee/AkkayaDEL16,smcworkshop}. 
These concerns are obviously not of the same kind and come from different stakeholders, requirements, or knowledge sources.

Let us consider a simple \textit{capture-the-flag} example~\cite{Mason2018,DBLP:conf/aaai/DeardenFR98}, in which a robotic system accommodates two autonomous software-driven agents constantly moving within an office building with connected rooms and hallways. 
The agents may communicate and collaborate to carry out the system mission --  collecting all flags scattered throughout the rooms. 
For such a robotic system, the complete model describing the different facets, such as the layout of the building,  the communication protocols, or  the behaviors of the robots,  would inevitably be quite complex. 
Validation would also be hard. 
For example, in order to determine a design solution that will achieve better system target satisfaction, one might want to explore the impact of different behavioral deployments, different communication protocols, or different spatial layouts before the actual system deployment. 
When these aspects are all intertwined in the global system model, it is difficult to explore and compare.

The principle of ``separation of concerns''~\cite{dijkstra1982role} is highly desirable in such situations, by which different system models are used to capture distinct concerns. 
That makes modeling and model-update simpler. 
Moreover, this principle allows model reuse that would be helpful for modeling of CPSS. 
In general, for CPSS, the space layout and the spatial activity are relatively unchangeable and may be reusable for different missions. 
However, the deployment of  system entities with specified behaviors and the interaction concerns among the system entities, may be dependent on the particular mission considered. 
On the other hand, system analysis and validation ask for the integration of these separate models so that it can be determined whether or not the overall system mission can be achieved. 
Whenever a change or substitution is made on an individual concern, the integration process needs to be repeated before analysis. Thus, systematic integration of models is a vital link in system model validation.  

To this end, this paper proposes a divide-and-conquer modeling methodology. 
That is, separate analyzable models capture recurrent concerns in CPSS, which are then systematically and semi-automatically formally integrated yielding automata equipped with transition guards, invariants and probabilistic features. 
Then, state-of-the-art statistical model checking techniques~\cite{bulychev2012uppaal,larsen2014statistical} are used for validation, prior to system implementation or deployment. 
Such validation methods may not offer definitive assurances like formal verification, but can provide valuable insights early in the design process and can scale to practical systems. 
This kind of feedback about  potential outcomes of early design choices in the development process can help the designer to explore the solution space and make decisions in a cost-effective manner. 
Our main contributions are summarized as follows:
\begin{itemize}
\item We propose a methodology identifying three key recurrent concerns representing CPSS facets compatible with existing formal modeling techniques;
\item We propose a semi-automatic method of integrating three models into an analyzable one capturing all concerns and a general algorithm of model integration, upon which the validation process is carried out;
\item We evaluate our approach over two different cases of CPSS, demonstrating its applicability.
\end{itemize}

The rest of the paper is structured as follows. Section~\ref{sec:relWork} summarizes some related work. Section~\ref{overview} presents an overview of our approach. Section~\ref{sec:concerns} presents separate modeling of different system concerns. Section~\ref{sec:comp} yields integrated model capturing conformable behaviors while Section~\ref{sec:smc} illustrates early requirements validation through statistical model checking. Experiments have been done to evaluate the approach in Section~\ref{sec:eval}. 
Section~\ref{sec:conclusions} concludes the paper.



\section{Related Work}
\label{sec:relWork}

This paper focuses on the modeling and validation of CPSS in the early design stages. 
Consequently, we classify related work into three categories. 
First, we discuss key approaches in CPS system design as multi-agent systems (MAS). 
Then, we review related techniques on engineering systems through integrating multiple concerns. 
Lastly, we discuss related CPS applications utilizing statistical reasoning, framing our approach within the overall software engineering domain.

Researchers have reflected that the metaphors of agents and the principles of multi-agent systems remain attractive for designing and engineering cyber-physical systems~\cite{DBLP:conf/emas/MascardiW18}, given the increasing integration and the inherent uncertainties CPS face. 
In this regard, there have been multiple efforts to design CPS using MAS engineering methods. 
In~\cite{DBLP:journals/tsmc/FortinoRSSZ18}, the agent-based computing paradigm has been explored to support IoT systems analysis, design, and implementation. Based on the agent-based cooperating smart object methodology and related middleware, effective agent design and programming models are provided along with efficient tools for the actual construction of an IoT system in terms of a multiagent system.
With applications to automated guided vehicles and transportation systems, an architecture-based design of MAS has been proposed that puts architecture at the center of the development activities by documenting specific concerns such as roles, organizations, and interaction protocols~\cite{Weyns2010}.  
In~\cite{DBLP:journals/cii/LeitaoCK16}, multi-agent systems have also been recognized as sharing common ground with CPSs and being able to empower CPSs with a multitude of capabilities, so to effectively enable emerging CPS challenges.
These works have demonstrated that the multi-agent paradigm has  potential advantages in CPS design, but to the best of our knowledge, they do not touch the point on how to effectively model and validate CPSs when problem complexity needs to be managed  and even complex system models need to be constantly adjusted and validated during the design phase.

Some approaches have recognized that the operating environment needs to be treated explicitly as a first-class abstraction in MAS which provides the surrounding conditions for agents to exist and an exploitable design abstraction~\cite{DBLP:journals/aamas/WeynsOO07}~\cite{DBLP:conf/e4mas/WeynsM14}~\cite{zhi2018Environment}.
The use of organizational concepts such as e.g., the AGR (Agent-Group-Role) organizational model has been adopted for describing the structures and the interactions that take place in MAS. 
In~\cite{DBLP:conf/e4mas/FerberMB04}, the AGR model is extended by assuming that agents are situated in domains, i.e., spaces, which may be physical (i.e. geometrical) or social. 
That allows to give a clear distinction between an agent and its mode, i.e. the way it appears and interacts into a space with other agents, aiming to show that a multiagent world is constituted of agents that may perceive and act in spaces and manifest their existence through their mode. 
This work explicitly uses the concept of physical space, but the space is not treated a first class abstraction and does not include its separate modeling. 
Some researchers propose the concept of intelligent virtual environments and develop an ontology comprising concepts for modelling intelligent virtual environments enhanced with concepts for describing agent-based organizational features~\cite{DBLP:journals/jaihc/DuricRCSJ19}.
The agents' environment has been proposed to be a first-class abstraction within MASs but it has not been modeled separately. 
In contrast, we explicitly model the spatial environment as a state-transition structure 
and consider the possible spatial behavior of active agents as a special concern. 

The separation of concerns is a cornerstone principle for complex systems since it can simplify development, maintenance and reusability~\cite{dijkstra1982role,DBLP:books/daglib/0014226}.
Aiming at handling the ``multiple perspective problem'' in composite systems in which there are multiple stakeholders involved, in~\cite{Framework_Multi_System_Development, MultiView_RequirementSpecification}, viewpoints are used to partition the system specification, the development method and the requirements representation. Using viewpoints to encapsulate the heterogeneous requirements from different stakeholders makes the requirements elicitation much easier. 
Such multiple-view conceptions have led to the interaction and integration of different viewpoints contributing to resulting requirements specifications~\cite{DBLP:journals/tse/NuseibehKF94}. 
Multi-view reasoning has also been adopted in architectures with  multiple and potentially conflicting concerns for quality requirements~\cite{MultiView_Architecture}.
Apart from requirements, complex software development must deal with more massive problem domain knowledge. 
By analyzing different models of object-oriented software development to identify the main differences in handling problem domain knowledge, a two-hemisphere modeling approach~\cite{DBLP:conf/caise/NikiforovaK04} has been put up to accommodate different models and to automate the process of  model transformation. 
It also enables knowledge representation in terms of business process models and concept models. 
Within the field of cyber-physical systems specifically, a conceptual model of a CPS has also been proposed to support 
different concerns (as views) of 
physical, cyber-physical, and computational aspects~\cite{tsigkanos2016architecting}. 
Those can be considered as the separation of different concerns. 
Our approach is similarly along this line. 
The referred concerns are three aspects about the construction of CPSS specifically, i.e. the physical space, the agent deployment and interaction, as well as the task requirements.

Similarly in the above mentioned practices of separation of concerns, building a comprehensive system model  benefits from or relies on model integration of the separate concerns or aspects~\cite{DBLP:journals/pieee/AkkayaDEL16}. 
There are many efforts on this topic. For example, the synthesis of behavioural models for modeling and reasoning about system behavior at the architectural level, such as labelled transition and modal transition systems, from scenario-based specifications has been extensively studied~\cite{DBLP:phd/ethos/Uchitel03,DBLP:journals/tse/SibayBUK13}. 
Previous work~\cite{fse17} has targeted automatically obtaining automata structures for cyber-physical spaces, which can bootstrap a core aspect of the present work -- spatial behaviour of entities from static space descriptions. 
Moreover, physical models themselves may be automatically obtained~\cite{bxcities}. 
Indeed, our approach is based on separate modeling of different concerns, systematic composition of separate models, and validation of the composite model. 
By supporting experts to focus on different models and then providing a way to integrate them in a systematic (and semi-automatic manner) -- instead of requiring a single model to be provided to encompass all views -- we can provide improved design support for CPSSs. 

Regarding validation, notable recent approaches have focused on applications of  statistical model checking in diverse domains within CPS, expanding the scope that is treatable beyond explicit-state verification~\cite{DBLP:conf/isola/LarsenL16,DBLP:conf/qrs/LiBPYJ18}.
In~\cite{ruijters2016better}, a framework utilizing statistical model checking for dependability within railway systems is developed. 
In addition, rare events problems in cyber-physical  applications have been emphasized in statistical model checking, either by adding feedback control to efficiently estimate  probabilities~\cite{DBLP:conf/isola/KalajdzicJLBLSG16} by importance sampling and Cross-Entropy methods~\cite{DBLP:conf/atva/ClarkeZ11}, or by importance splitting and reformulating rare probabilities~\cite{DBLP:conf/cav/JegourelLS13}.
Also within software engineering, the ActivFORMS~\cite{deltaiot} framework exploits statistical model checking at runtime to select configurations that comply with self-adaptation goals over an internet-of-things network topology. 
In contrast, our approach targets early requirements validation through separation of system design concerns.




\section{Approach Overview}
\label{overview}





Design of cyber-physical space systems must take into account their spatial environment and how system requirements can be affected by the behavior of various active agents. 
The specific spatial environment  a system is found in, dominates agent behavior -- it delineates the spatial actions that are possible within it.
Besides actions in space, agents in a cyber-physical space system also interact. Typically, this interaction may take the form of communication or coordination, as agents do not operate in isolation but are part of a composite system.
Thus, the overall system exhibits composite behaviors which may satisfy or violate its design requirements.
Our approach concerns high-level reasoning during the system requirements phase, where a way to validate system behaviors before implementation and deployment is highly desirable to provide the required assurances for the final system.
The main driver of our approach is separation of concerns -- the design principle for separating a design into distinct sections, such that each section addresses a separate concern~\cite{dijkstra1982role}.

\begin{figure*}[htbp]
  \centering
\includegraphics[width=15cm]{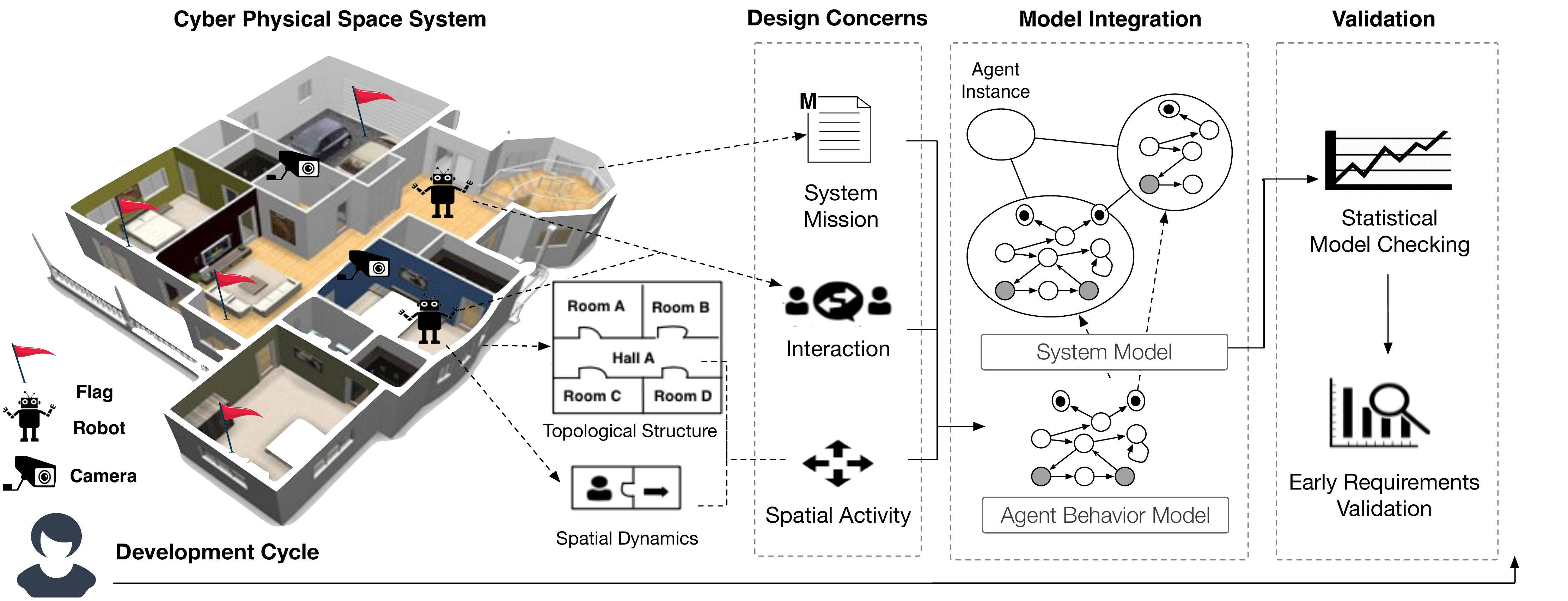}
\caption{
System Modeling, Model Integration and Validation for CPSS:
Approach Overview.
}
\label{approach}
\end{figure*}

Figure~\ref{approach} shows a bird eye's view of our approach. 
The development cycle of a cyber-physical space system starts by taking into account the spatial environment and the possible spatial behavior of active agents (from left to right in Figure~\ref{approach}).
This is because a CPSS is usually coupled by multiple \emph{problem solvers}, i.e., agents with certain capabilities which operate to achieve some mission. 
Mission achievement can be specified by describing certain desirable states that those agents are required to bring about~\cite{DBLP:journals/aamas/WeynsOO07}. 
Additionally, active agents in the CPSS may communicate and coordinate with each other in order to collectively achieve the system mission. 
Thus, we identify three distinct concerns: (i) \emph{spatial activity}, (ii) \emph{interaction} and the overall (iii) \emph{system mission}.
Within our approach, those are captured \emph{independently} by following a well-defined and rigorous modelling method  (Section~\ref{sec:concerns}). 
Models are then semi-automatically integrated leading to coherent behavioral model (Section~\ref{sec:comp}), 
capturing behaviors of a single autonomous agent.
 Then, the system model can be obtained by considering the collective of interacting agent instances. 
Analysis of the system's mission achievement can then be performed by validating the system models -- in our approach we advocate requirements validation  through statistical model checking. The analysis results finally acquired can guide implementation and deployment of the actual system modeled.

\noindent
\textbf{Running Example.} 
As a motivating example showcasing our approach, consider a \emph{capture the flag} mission~\cite{Mason2018} often used as a benchmark for mission planning~\cite{DBLP:conf/aaai/DeardenFR98} in artificial intelligence or robotics domains. In such a mission, flags as static physical objects are scattered throughout a building comprising connected rooms and hallways. A team of two active robots are dispatched to find these flags. The objective of this team is to achieve collection of all the flags. However, the building is augmented with security cameras that monitor certain areas; a camera scans a designated area, surveying for possible intruders. The agents searching for the flags do not know beforehand the location of surveillance cameras. Detection of an agent twice by the camera results in the capture of the agent and its termination. Agents may attempt to communicate the position of a security camera an agent has located, to the other. If communication is successful, the other agents avoid entering the corresponding location.



For the purposes of this motivation example, we are not concerned with planning but with validation\footnote{Behaviour of agents adhering to a strategy can be additionally encoded.}. A system goal concerns the composite CPSS that the active agents and cameras induce, and entails that
\emph{all three flags must be collected within 10 time units while no more than one robot agent must be terminated}. A time unit is defined as the execution of one spatial movement by a robot, such as changing its location to another room.

\section{Modeling Concerns in CPSS}
\label{sec:concerns}


In the following, we describe a systematic way to model the three major concerns in CPSS. 
We begin by presenting our formalism of choice, which aims at capturing in a precise manner models of these concerns sourced from appropriate domain models. Thereupon, we show how spatial activity and interaction as well as the system missions can be modeled. 


\subsection{Modeling Formalism}

Stochastic Timed Automata (STA~\cite{rodriguez2013using}) are one of the prominent classical formalisms for describing behaviors of real-time systems~\cite{baier2008principles} such as ones consisting of cyber-physical components and stochastic features~\cite{Beauquier03}. Our choice of STA is motivated by the fact that it is generic enough to encompass various domain models describing spatial behavior, while enjoying precise integration semantics necessary for requirements reasoning. 
Moreover, uncertainty is common in CPSS and agents' behaviour is often probabilistic and time-sensitive.
We begin by describing timed automata briefly, then we stepwise enrich the model including stochastic aspects. 
The interested reader can refer to foundational works~\cite{alur1994theory,rodriguez2013using,baier2008principles,Beauquier03} for complete definitions and precise semantics.


A \emph{timed automaton} (TA~\cite{alur1994theory}) is a tuple $\mathtt{TA = (Q, q0, X, G, T, A, Z)}$. 

\begin{itemize}
\item $\mathtt{Q}$ is a finite set of states;
\item $\mathtt{q0} \in \mathtt{Q}$ is an initial state;  
\item $\mathtt{X}$ is a finite set of real-valued variables called clocks;  
\item $\mathtt{T}$ is a set of transitions;
\item $\mathtt{G}$, a set of \emph{guards}, control the triggering of transitions from state to state during an execution;
\item $\mathtt{A}$ is the set of \emph{actions} attached to transitions;
\item $\mathtt{Z}$ are the invariants assigned to individual states and expressing further constraints on delay times in clocks.
\end{itemize}

A guard is an expression with no side-effects which evaluates to a boolean value and is attached to every transition; an expression value of \emph{true} will enable the transition for choice, while it will disable it if \emph{false}. Actions set $A$ is partitioned into \emph{input} ($I$), \emph{output} ($O$) or \emph{internal} ($\Gamma$) actions. Finally, a transition $T$ of a $\emph{TA}$ can be specified as a tuple $\mathtt{t = (q, a, g, q')}$, which specifies a transition from state $\mathtt{q}$ to $\mathtt{q'}$ with actions $\mathtt{a}$ (either input, output, or internal action) and guards $\mathtt{g}$, where $\mathtt{q,q' \in Q; a \subseteq I \cup O \cup\ }$ $\Gamma$  and $\mathtt{ g \subseteq G}$. 

To capture stochastic behaviors in a timed automaton -- yielding a stochastic timed automaton (STA~\cite{rodriguez2013using}), two aspects are introduced: states of the $\emph{TA}$ are associated with a probability density function ($\mu$) and sets of transitions (from each state) are associated with probabilistic choices. 
Time delay over a state is not fixed but follows the distribution (e.g., exponential distributions) according to $\mu$. 
Observe that a state might have multiple successor states, each having guards. Informally, guard semantics is as follows. First, guard expressions are evaluated, enabling or disabling sets of transitions exiting the state. Among the enabled sets of transitions, a non-deterministic choice selects a set of transitions. From that set of transitions, one is chosen depending on the defined probability distribution over the successor states~\cite{NormanPS13}. 
Specifically, if $\mathtt{T_{qg}}$ is the non-empty set of transitions starting from $\mathtt{q}$ with the same guard $\mathtt{g}$, then for all $\mathtt{q \in Q }$, it holds that  $\mathtt{ \sum_{t \in T_{qg}} \rho(q,t) = 1}$. 
Transitions of an $\emph{STA}$ capture the following behavior. When a state is entered, a wait time is chosen; after the wait time has passed, a transition is enacted according to the defined transition probability~\cite{rodriguez2013using}. The STA we consider are supported by \emph{$\mathtt{uppaal-smc}$}~\cite{bulychev2012uppaal}, while further effective  tool support widely exists for analyses based on such STA~\cite{david2011statistical,prism}.

\subsection{Modeling Spatial Activity}
\label{physicalconcern}


The rationale of conceiving separately a spatial activity model in our approach is that spatial behavior of agents within an environment can be derived from domain information. Domain information can encode how the space is structured in a topological manner as well as how active agents change their location within this topological structure -- essentially, agents change their discrete location within the spatial environment. This is an adequately general model that can encompass various others. 
Such a spatial activity model typically has a form of a state-transition structure, where states represent possible locations that autonomous agents may be found in, and transitions represent how they may move from a location to another. The model can either be constructed manually or automatically derived by sourcing appropriate domain representations. Such domain descriptions can be modeled using e.g., process calculi yielding state-transition structures~\cite{Foster06adaptablesoftware,fse17} automatically. 
Another way of obtaining such a model is based on domain descriptions representing spatial space~\cite{bxcities}, such as Building Information Models (BIM~\cite{BIM}) or CityGML~\cite{kolbe2005citygml}), upon which dynamics of autonomous agents are encoded via transformation rules~\cite{tsigkanos2016architecting}. Furthermore, domain models of geolocation trajectories of e.g., internet-of-things devices can be used~\cite{amelia}.

\begin{figure}
  \begin{subfigure}[b]{0.4\textwidth}
\includegraphics[width=\textwidth]{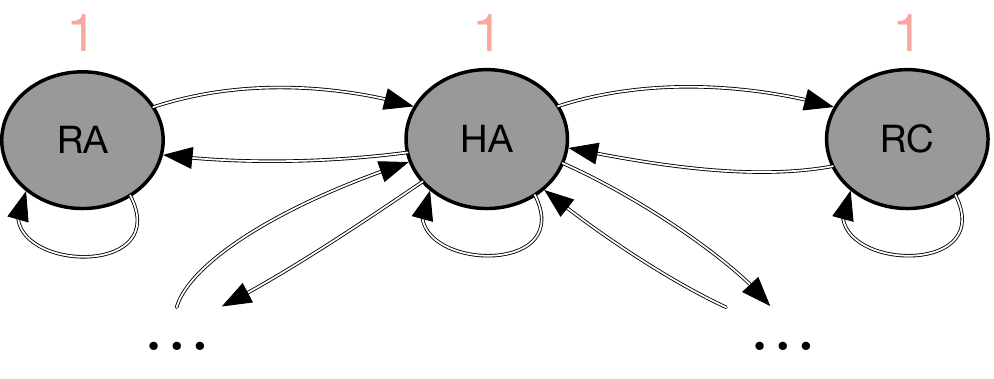}
\caption{Spatial Model}
\label{physicalactivity}
  \end{subfigure}
  \hspace{1.1cm}
  \begin{subfigure}[b]{0.4\textwidth}
  \centering
\raisebox{1cm}{\includegraphics[width=\textwidth]{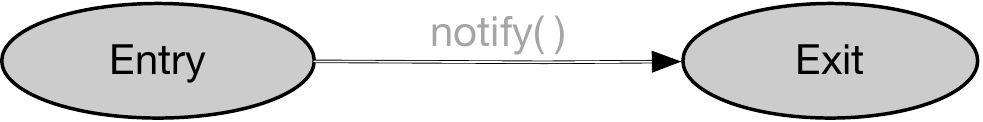}}
\caption{Interaction Model.} 
\label{uavinteractionsimple}
  \end{subfigure}
  \caption{Activity models for the capture-the-flag  example; a robot may move between certain physical locations (a), and independently interact with other robots (b) through a notify() operation.  }
\end{figure}

Recall the motivation example;
the spatial activity model of robots is represented by an STA which can be derived with the aforementioned techniques, such as the structure of the building~\cite{sescps}.
 A state records the position of an agent in some location through a proposition. For example, a robot agent might be located in room A, represented with the state labeled \emph{RA} in the STA of Figure~\ref{physicalactivity}. Transitions reflect possible change of location to another. For example, if an agent is in a hallway named \emph{HA}, and adjacent to it there is a room named \emph{RA} and another room named \emph{RC} with connecting doors, the STA has a transition from a state labeled \emph{HA} to one labeled \emph{RA} and to one labeled \emph{RC}. 


The model obtained can be further enriched with additional domain information. If the time spent in a location is known, this can be specified and incorporated to the model accordingly. This may be sourced e.g., from knowledge about the geometrical size of the rooms in a building, which require the robot to stay a longer time searching for the flag.
Instead of explicitly defining clock variables in the STA thus encoding directly time units invariant in each state, stochastic behavior of the agent is specified -- we assume that time units in every state of the STA are obtained from some probability distribution with rates supplied by the system designer. For our motivating example and the robot activity of Figure ~\ref{physicalactivity}, the robot agent may stay in Room A $RA$ for an average time of 1 time unit, e.g. in an exponential Poisson distribution, in the absence of invariant $Z$ to the states~\cite{UPPAALsmcTutorial}.

\subsection{Modeling Success and Failure}


In general, system missions may refer to cross-cutting system concerns, and may refer to either quantitative or qualitative objectives of a set of agents. In our motivating example, recall that the system goal states that the ``all four flags must be collected within 10 time units while no more than one robot agent must be terminated''. 
A set of elementary predicates about one or multiple agents in the cyber-physical space system, which are composed in a logical manner that introduces quantitative or logic constraints about single agents, could then be manually derived. For instance, the fact that an agent is in a specific room or is in a condition of termination due to successful monitoring by the surveillance camera are predicates for a single robot agent. Each predicate may be either true or false for an agent. 


Thus, states in an automaton expressing behavior can represent success or failure of predicates. In our setting,
to enable reasoning such predicates are encoded as success-failure states in an STA capturing each agent; if an STA is found in such a state, the predicate is considered satisfied (or violated) for that particular agent modeled by the STA. 
This designates \emph{reachability properties} as the fundamental requirements building block, as success-failure states reflect that some goal or failure of an agent has been reached. 
In our capture-the-flag example, a ``terminated'' failure state is introduced, which reflects the robot agent's status as terminated. Such a state is absorbing -- the STA modelling the agent should not be able to continue operating. 
These success-failure states along with auxiliary primitives and system variables become available for the overall system requirements specification (illustrated later in Section~\ref{sec:smc}).
\subsection{Modeling Interaction}


In a CPSS, agents do not operate in isolation; they also interact. Through interaction, they may also coordinate behaviors. Automata are frequently used to model interaction, such as one occurring between components in a system~\cite{componentinteractionautomaton}, between humans and robots~\cite{humanrobotinteraction}, and protocols among agents in multi-agent systems~\cite{multiagentprotocolautomaton}. 
Thus, STAs -- being quite expressive, general automata -- can be naturally used to express agent interaction and coordination as well.
Communication between agents operating in a CPSS for instance, is a typical form of interaction, and a model of a communication protocol describes how interaction between agent instances takes place.
The interaction concern between agents may differ in different scenarios; faithful to the separation of concerns principle, we describe the interaction model via a separate STA.
Typically, the STA describing the interaction model may be specified independently by a domain expert; e.g., an expert in communication protocols. The interaction between different classes of agents is also allowed.

To ensure that various aspects of our approach are compatible and conformable, we assume that the interaction model is sourced from some domain model and it is an STA $\mathtt{IT = (Q, q_1, q_n, X,}$ $\mathtt{ G, A, T, Z)}$ 
with both an initial and a final state, where the initial state $q_1$ signifies the start of the interaction logic and the final (exit) state $q_n$ signifies the end of the interaction logic.
Given transfer of control to the entry state, progression to the exit state has to always eventually occur -- in other words, the interaction or communication protocol must be always terminating.




Figure~\ref{uavinteractionsimple} illustrates the rudimentary case of communication between robot agents in the example capture-the-flag mission. 
For this small example, interaction is simple and consists of a single one-way operation (\emph{notify}).
For our robot scenario, time-units spent in each state of interaction automata are considered negligible for simplicity. A set of shared variables common to the various agents can be used in conjunction with transition guards or actions by the system designer to implement interaction logic. 
Note that in real systems interaction and communication  models can be very complex (such cases will be illustrated later in Sec.~\ref{sec:eval}).

\section{Model Integration of Agent Behavior}
\label{sec:comp}

While the aforementioned models capture separate concerns, a unified model capturing spatial and interaction activities and success-failure states of autonomous agents is needed. We illustrate in this section how such models can be integrated, enabling requirements reasoning on the overall CPSS induced by its agents.

The key intuition behind incorporating success-failure states is that an agent may be at any point found in a situation that fulfills an elementary predicate; in other words, the STA capturing the agent's behavior may enter a success-failure state which encodes such a predicate. 
Similarly to incorporating success-failure states, the composition of interaction with the spatial activity is performed for every of the spatial activity STA's states; that is to say, possible interaction could occur from any spatial location that an agent may be found in.
Triggering interaction may be of course context-dependent, something which can be specified explicitly by the designer by encoding appropriate guards. Such a guard would predicate on conditions that would enable or disable interaction based on some context that is true when the agent is in the relevant spatial STA state. We consider by default that guards controlling transitions to interaction STA are checked (and a transition to interaction STA executed) before transitions to other spatial STA states are considered by the integrated automaton. 
To precisely define the composite agent behavior, let $R = (sf1, sf2,...,sfn)$ be the set of success-failure states of an agent, $P = (Q_{P}, q0_{P}, X_{P}, G_{P}, A_{P}, T_{P}, Z_{P})$ its spatial activity automaton, and ${I = (Q_{I}, q1_{I}, qn_{I}, X_{I}, G_{I}, A_{I}, T_{I}, Z_{I})}$ be an interaction STA. We assume that $Q_{P} \cap R = \emptyset$. The automaton ${C =(Q_{C}, q0_{C},  X_{C}, G_{C}, A_{C}, T_{C}, Z_{C})}$ describing the integrated agent behavior and consisting of all three concerns has the following form:

\begin{itemize}
\item $Q_{C} = Q_{P} \bigcup R \bigcup Q_{I}$ where $Q_{P}$, $R$,  $Q_{I}$ are disjoint;
\item  $q0_{C} = q0_{P}$;
\item  $X_{C} = X_{P} \bigcup X_{I}$;
\item $G_{C} = G_{P} \bigcup G_{I} \\ \hspace*{1.5cm}\bigcup \{ check\_sf_i == true, \\\hspace*{2.0cm}check\_sf_i == false\ |\ sf_i\in R\}\\\hspace*{1.5cm} \bigcup \{ check\_interaction == true,\\\hspace*{2.0cm} check\_interaction == false \}$\\/* two new added guards for each predicate state and two for interaction*/;
\item $A_{C} = A_{P} \bigcup A_{I}\\\hspace*{1.5cm}\bigcup \{action\_spatialactivity \} \\\hspace*{1.5cm}\bigcup \{action\_sf_i\ |\ sf_i\in R\}
\\\hspace*{1.5cm} \bigcup \{action\_{q1_I}, action\_{qn_I}\}$
\\/*added actions encoding any additional logic for spatial activity, predicate states, and before entering and after exiting the interaction*/;
\item $T_{C} = $
$\\\hspace*{0.4cm}\{ (q, a, g, q')~| 
	\\\hspace*{1.3cm}q \in Q_{P}, q' \in R, \\\hspace*{1.3cm}a = \{ action\_{q'}\},\\\hspace*{1.3cm} g = \{ check\_{q'}  == true \} \} \\  
\bigcup $
$ \{(q, a, g, q')~|\\\hspace*{1.3cm}q \in Q_{P}, q' = q1_{I},\\\hspace*{1.3cm} a = \{ action\_{q1_I} \}, \\\hspace*{1.3cm}g = \{check\_interaction == true\}\\\hspace*{1.6cm}\bigcup\{ check\_{sf_i} == false\ |\ sf_i\in R \} \}\\ 
\bigcup \{ (q, a, g, q')~| \\\hspace*{1.3cm}q = qn_{I}, q' \in Q_{P}, \\\hspace*{1.3cm}a = \{ action\_{qn_I} \}, \\\hspace*{1.3cm}g = \{\} \} \\ 
\bigcup  \{ (q, a', g', q')~| \\\hspace*{1.3cm}(q, a, g, q') \in T_{P} ,\\\hspace*{1.3cm} a'  = a \bigcup \{ action\_spatialactivity\}, \\\hspace*{1.3cm}g' = g \\\hspace*{1.6cm}\bigcup \{ check\_{sf_i} == false\ |\ sf_i \in R \} \\\hspace*{1.6cm} \bigcup \{ check\_interaction == false \} \} \\
\bigcup T_{I}$.
\item $Z_{C} = Z_{P} \cup Z_{I}$.
\end{itemize}

Transitions are added from every state of $P$, accounting for each success-failure state, interaction start and interaction end yielding control back to $P$. This is to ensure that the agent might potentially reach its absorbing state or interact from any spatial location. Transition guards define and evaluate a condition that may lead (or not)  to certain states.  Conditions $\mathtt{check\_{sf_i}()==true}$ are attached to the transition from the spatial STA's states to each $sf_i$ in success-failure states, $\mathtt{check\_{sf_i}()==false}$ and  $\mathtt{check\_interaction()==true}$  between the spatial STA's states to interaction entry states, and $\mathtt{check\_{sf_i}()==false}$ and  $\mathtt{check\_interaction()==false}$ between spatial activity states guarantee the high-level control flow of the integrated STA. 
These transition guards ensure that high-level control flow is maintained: (i) success or failure is always checked first for the agent, (ii) interaction is subsequently attempted and finally (iii) spatial movement occurs, while (iv) real-time does not pass when transitioning between automata that capture different concerns.

Function prototypes corresponding to application-specific logic are further attached automatically 
to relevant transitions as internal automata events, and are available for 
implementation based on the overall system application. For example, $\mathtt{action\_sf_i}$, $\mathtt{action\_q1_I}$, $\mathtt{action\_qn_I}$ and $\mathtt{action\_spatialactivity}$ are added -- their implementation is left to the system designer, who can utilize domain knowledge to implement concerns that e.g., span different (classes of) agents. 

\begin{figure}[h]
  \centering
\includegraphics[width=10cm]{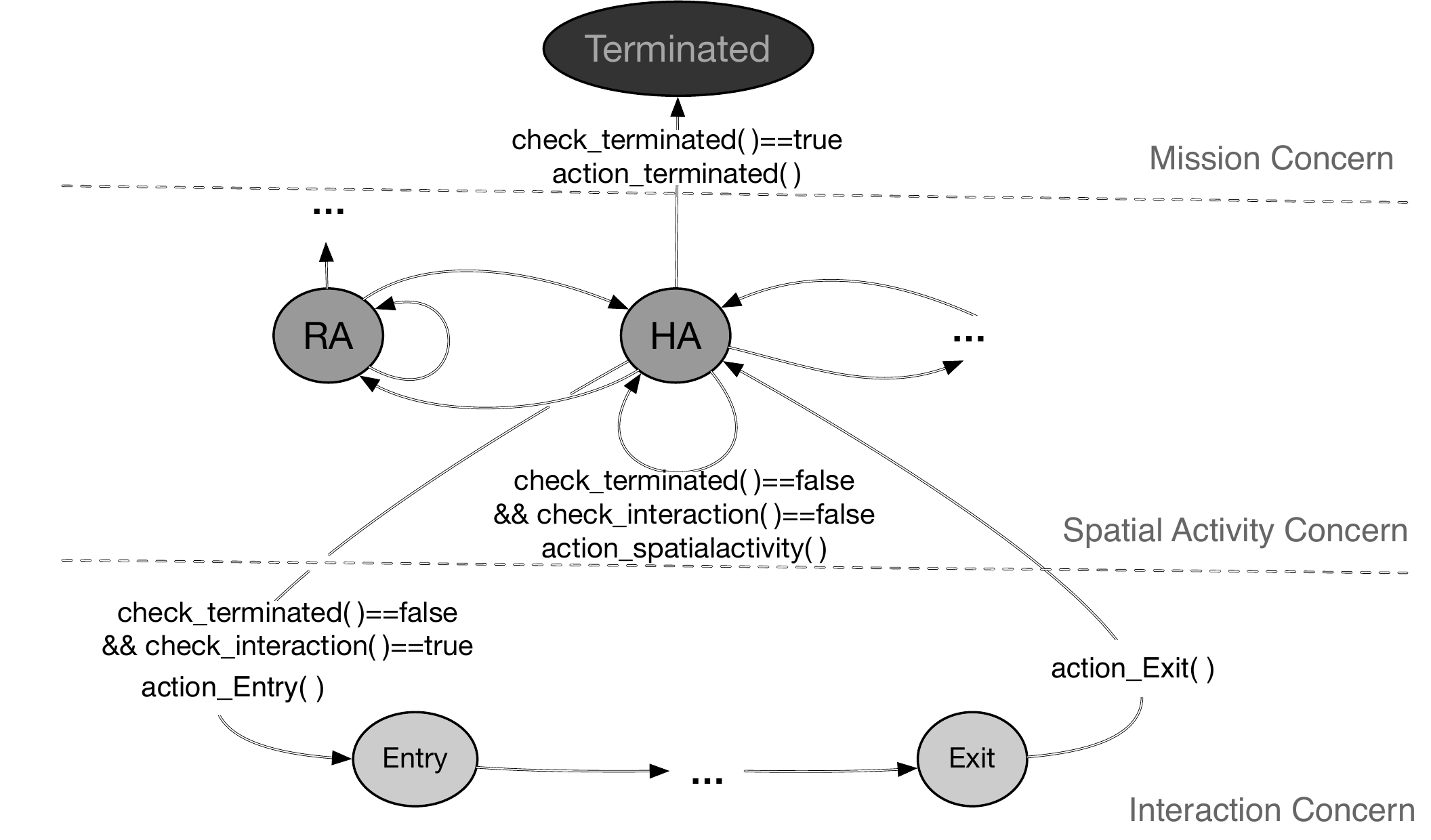}
\caption{Fragment of the integrated robot STA.}
\label{agentautomaton}
\end{figure}

In our approach, implementation of function prototypes  
in transition guards and actions is delegated to the system designer, who can utilize domain knowledge to encode predicates inherent in the system missions. This renders our approach semi-automatic. 
By delegating this encoding 
to the designer, a variety of domain-specific behaviors can be modeled, utilizing available primitives such as spatial locations and current agent positions.

\begin{lstlisting}[basicstyle=\footnotesize, caption={Partial Implementation required for the motivating example.},language = C,label=funcimplementation]
bool check_terminated(){
	if(detected_time==2) return true;
	else return false;
}
void action_terminated(){
	robot_pos = -1;  
}
void action_spatialactivity(){ 
	if(camera[robot_pos]==true)  
  		detected_time++;  
}
\end{lstlisting}

Back to our motivating example, a fragment of the agent model generated is illustrated in Figure~\ref{agentautomaton}. In the middle part, the spatial activity captures position of an agent in locations $RA$ or $HA$. From location $HA$, the STA may enter either its failure state, or the interaction STA (lower part), where the robot may notify the other one of its local information (camera or not). 
For a robot agent, implementation of guards and actions is shown in Listing~\ref{funcimplementation}. The function $\mathtt{check\_terminated()}$ encodes the case where a robot has been detected twice. If this is the case, 
function $\mathtt{action\_terminated()}$ encodes that the robot position information is erased, triggered as the agent is terminated. When the robot moves to a new location, the time of detection will be set to one if there is a camera inside the room, as shown in function $\mathtt{action\_spatialactivity()}$.
Note that function prototypes $\mathtt{action\_spatialactivity()}$ may encode other arbitrary logic for the agent, to be executed when changing spatial locations. 
The complete guard and function prototypes implementation is omitted for clarity in Figure~\ref{agentautomaton}; the complete specification of the motivating example can be found in accompanying material~\cite{paperstuff}. 
\section{System Validation}
\label{sec:smc}




In this section, we discuss how requirements of the CPSS are specified, and subsequently give an overview of how statistical analysis of their satisfaction upon the system is performed. 
Given a CPSS mission, the system designer specifies i) requirements of interest as logical formal properties and the ii) deployment setup, upon which analysis will take place. The latter entails parametrization of the CPSS depending on deployment of the system under investigation -- the deployment setup submitted for analysis specifies the number of autonomous agents and their initial states. The deployment configuration is subsequently loaded in the off-the-shelf statistical model checking tool \emph{$\mathtt{uppaal-smc}$}~\cite{UPPAALsmcTutorial}, and results are obtained.





\subsection{Property Specification}

Recall our capture-the-flag example; 
the system mission concerns the whole system that the robot agents induce. Firstly, it predicates about certain robot(s) having some property, thus reasoning about multiple agents is required. Secondly, it predicates about their overall behaviour in the  space (i.e. the number of flags they collect). Finally, it predicates about the passage of time within the mission. 

To investigate if the CPSS satisfies a requirement, it must be expressed in a manner that analysis can be enabled. To this end, we adopt Metric Temporal Logic~\cite{DBLP:journals/rts/Koymans90}, -- a timed extension of Linear Temporal Logic~\cite{clarke1999model} -- expressing properties over execution runs of the system, defined as:

\begin{eqnarray} 
\phi ::= \text{ap} ~|~ \neg \phi ~|~ \phi_1 \wedge \phi_2 ~|~ \phi_1 \vee \phi_2 ~|~ \bigcirc \phi ~|~ \phi_1\ \cup_{x \leq d }  \phi_2. \nonumber
\end{eqnarray} 

\noindent
In the grammar, $ap$ is a proposition,
$d$ is a non-negative integer and $x$ is a clock. The logical operators are interpreted as usual, and $\bigcirc$ is a next state operator. $\phi_1\ \cup_{x \leq d}  \phi_2$ is satisfied by a run if $\phi_1$ is satisfied on the run until $\phi_2$ is satisfied, and this will happen before the value of the clock $x$ exceeds $d$. It provides additional reasoning upon clock variables and clock constraints that specify timing behaviours. Given fundamental operators $\bigcirc$ ("next") and $\mathtt{U}$ ("until"), we can derive additional ones such as $\diamond \phi = true\ \mathtt{U}\ \phi$ ("eventually") and $\Box \phi = \neg \diamond\neg \phi$ ("always"). Thereupon, we define $P(\phi)$ to be the probability that a random run of the system satisfies $\phi$.

CPSS property specification using the syntax described occurs by utilizing four types of available propositions:

\begin{enumerate}


\item Spatial locations within the space where an agent may be found during execution. For example, the proposition $[RobotA.RC]$ in the capture-the-flag scenario reflects the fact that the robot $A$ is in the room $C$.

\item Agent success/failure of agents. For instance, the proposition \emph{RobotA.terminated} (resp. \emph{RobotA.successful}) expresses the fact that during an execution, robot $A$ is terminated (resp. successful) -- its behavior reached the relevant absorbing state. 

\item Auxiliary propositions that regard i) counts of agents and ii) time that they spend in specific spatial locations. For example, the (boolean) proposition $[RobotSFNum<=1]$ encodes the fact that the total number of robots that jump into their respective success-failure state is less than one, and $[SystemTime<=10]$ specifies that the time units spent in the system are no more than ten.

\item Application-specific global variables. The system designer is allowed to define custom global variables within function implementation, where those variables are exposed as propositions.
For instance, a global variable $[nFlag]$ maintains the count of flags collected by the robots; boolean propositions are derived from them such as $[nFlag == 3]$, representing the fact that the count of collected flags is 3.

\end{enumerate}

\noindent
Given the above types of propositions available, the capture-the-flag mission requirement can be expressed in the following formula, which states that eventually, in the system's execution after a maximum of 10 time units, the number of flags collected is three, and the count of robots reaching success-failure is at most one:

\vspace{-0.3cm}
\begin{eqnarray} 
\diamond_{[SystemTime<=10]} [numFlag==3] \wedge [robotSFNum<=1]. \nonumber
\label{flagprop}
\end{eqnarray}

\subsection{Early Validation  with Statistical Model Checking}

Statistical model checking (SMC)~\cite{younes2005verification,legay2010statistical} is a method for calculating the likelihood of the occurrence of certain events during execution of a system. This is performed through simulation runs, reaching some confidence level. 
Statistical techniques for analysis have been found to be applicable for large and complex systems that cannot be verified with classical model checking~\cite{UPPAALsmcTutorial}. 

To apply statistical model checking techniques, what is essentially required is i) a formal model describing a system able to generate finite sets of executions serving the purpose of observations, ii) a monitoring procedure to decide whether an execution satisfies the property under consideration and iii) a statistical algorithm yielding overall results for the system. 
The system model is used to generate execution traces upon which statistical methods produce statistical evidence about the system's satisfaction or violation of a property specification. 
In essence, for all available behaviors of agents in the systems at every moment, each simulation run picks up one path stochastically and returns ``true'' or ``false'', indicating whether or not the model of a system satisfies the system property for that run.
Subsequently, the designer obtains results useful to early requirements validation of the overall system~\cite{bohlender2014review}.

Recall that the integrated behavior model outlined in the previous sections represents one class of active agents -- in a CPSS, there would be various agent models concerning different classes. 
We instantiate as many instances of these classes depending on the CPSS and deployment configuration specified by the designer. The result is a network of STAs. Initial states and the number of agent instances depend on the deployment evaluated by the system designer.
The system model, along with a specified deployment configuration and property are subsequently loaded in the off-the-shelf statistical model checking tool \emph{$\mathtt{uppaal-smc}$}~\cite{bulychev2012uppaal} and analysis of the STA is invoked. The result is the degree of satisfaction or violation of properties with some obtained confidence level. Generally speaking, confidence represents the intervals that contain the true value of result from an infinite number of independent statistical samples. 



We succinctly indicate results of analyses of the capture-the-flag scenario of Section~\ref{overview}. 
For this scenario, two instances of the robot agent model are deployed with initial states in room $A$ and $C$ respectively. 
The configuration (including the initial position of robots, flags and cameras) is as shown in Figure~\ref{approach}. 
The 95\% confidence interval, containing the range of potential values of achieving the system goal, is [0.553, 0.652] within an average of 8 time units passing.  


\section{Evaluation}
\label{sec:eval}

To evaluate our approach and assess its applicability for validation, we consider two cases of spatially-dependent component systems.
The systems are different in domain, complexity, size and analyses required.
However, they both are within settings where different classes of active agents operate in a space-dependent environment.
The first models a swarm robotics system obtained from literature~\cite{honeybees}, while the second concerns a complex case of emergency response with autonomous Unmanned Aerial Vehicles (UAVs). The swarm robotics case intends to reproduce the original system and illustrate its stability and scaling attributes. The emergency response case tackles typical design questions within such a scenario, where high complexity is prevalent both in the system, agent interaction and spatial domain.

To concretely support evaluation, we realized a proof-of-concept implementation, which is available as an open source tool~\cite{paperstuff}
reflecting the integration and specification procedures of Section~\ref{sec:comp}.
 Models produced representing various classes of agents are compatible to \emph{$\mathtt{uppaal-smc}$}~\cite{bulychev2012uppaal} with which statistical validation is performed. 

\subsection{Honeybee Swarm Micro-Robotics}

For this evaluation case, we closely model a micro-robotics system~\cite{honeybees} from the robotics literature. Our objective is to perform a study of authors' findings through our model-based approach. 
The robotic design in question concerns a collective of resource-constrained micro-robots which move autonomously within a plane, each equipped with a sensor. The overall goal of the collective is to assemble at the location with the highest sensing value.
Such a swarm behavior is based on the aggregation behavior observed in honeybees, which aggregate at the warmest spot on the comb. Specifically, the bio-inspired micro-robots operate as  follows:

\begin{itemize}
    \item Robots move randomly within the spatial plane;
\item Robots detect collisions with others. If they collide, they stop.
\item A sensing measurement is taken  only when robots stop after a collision.
The higher the measurement, the longer they remain in the same location.
\end{itemize}

We note that the system has particular swarm characteristics; 
there is no communication involved, robots have no memory, and the swarm algorithm works also with poor sensor reliability. Moreover, there is no global knowledge needed for the swarm system's operation. 
In essence, a single robot has no chance to find the optimum, since it does not have memory and does not collide. As the number of robots  increases,  collisions increase as well, so the emergent collective behavior is successful in finding the optimum.
The main hypothesis we seek to investigate is the following~\cite{honeybees}:

\begin{displayquote}
``High swarm densities lead to more collisions, higher frequencies of sensing and thus faster convergence''.
\end{displayquote}



According to the principles mentioned in~\cite{swarm_intelligence}, the collective behaviors of the micro-robot swarm would show the following properties: i) stability, where the swarm should find a stable final solution whatever the initial distribution, and ii) scalability, where the algorithm should work better with greater numbers of robots. 
We consider temperature as the sensor measurement and reproduce the robotic setup described in~\cite{honeybees}. In the following, we briefly describe the robot modelling activities. Specifications and models are available in the online appendix~\cite{paperstuff}.


\emph{Agents' Physical Activity.} To generate the model of the micro-robot spatial activity, the area is divided into a grid. Each cell represents one state where the robot is located at a time, and neighboring cells denote the state-transition structures representing spatial moves in four possible directions (north, south, west, and east). The temperature for each grid and the possible resting time in a specific temperature for a robot are specified in advance as domain knowledge. The initial swarm distribution is chosen randomly to experimentally evaluate stability.

\emph{Agents' Success and Failure.}
We identify a success state for a robot denoting the successful finding of the "Optimal Cluster" within the spatial plane, where the temperature is the highest.

\emph{Agents' Interaction and Coordination.}
In the scenario studied~\cite{honeybees}, robots identify collisions with passive objects and other robots by emitting short-range light signals.
We simplify this behavior by ignoring passive objects and limiting interaction range within the grid. 
A micro-robot model integrating the aforementioned concerns 
is constructed with our framework; the designer may further specify number of micro-robot instances as well as initial states for each. 
For this experiment setup, 20 robot agents are initially deployed in a temperature field discretized into 10x10 grids. The temperature in the arena ranges from 22$^\circ$ to 36$^\circ$ (i.e., optimal) with the resting time in each grid from 1 to 6 time units. We study the variation in analysis results introduced by 1) the initial distribution of micro-robots to investigate stability, and 2) the number of robots to verify the property of scalability. The property specification we consider is $\diamond_{[SystemTime<=1000]} OptimalClusterNum >= threshold$. Here, we define a solution of cluster (i.e., threshold) as more than two-thirds of the robots gathering when total the number is no more than 15, and as 15 in other cases.

\paragraph{Stability of Swarm Behavior.}
We consider several different initial distributions for 20 robots.
From the experiment results, we observe that a swarm size of 20 is enough to form the optimal cluster no matter the initial configuration, though with different timings. Therefore, we investigate the time units needed to form the optimal cluster. Firstly, three groups of initial randomly scattered robot distributions are selected. The time to form the optimal cluster is around 1100 to 1200 time units, with probability beyond 90\%. 
Subsequently,
we investigate the case where all robots are deployed in one grid, a non-optimal cluster at very first beginning. The time for the cluster changing from 30$^\circ$ to 36$^\circ$ is around 1350 while the worst case from 22$^\circ$ takes more than 1600 time units. Overall, the swarm can find a stable optimal cluster independent on the initial distributions in the temperature field.

\begin{figure}[htbp]
  \centering
\includegraphics[width=10cm]{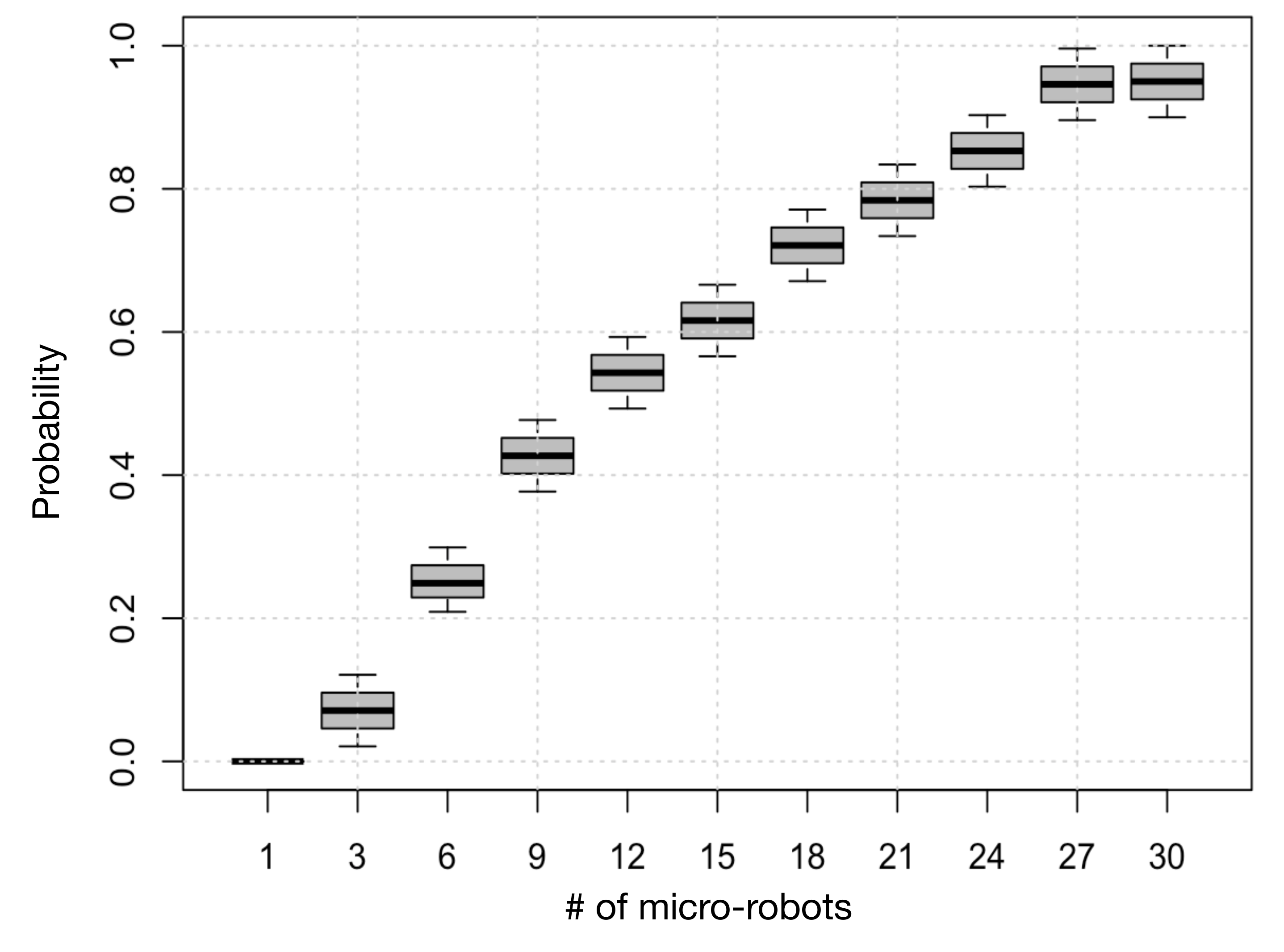}
\caption{Scalability over increasing number of micro-robots.
}
\label{fig:bee_scalability}
\end{figure}

\paragraph{Scalability of Swarm Behavior.}
For this set of experiments, we increase the number of robots in a stepwise manner and study its effect on optimum cluster finding, with respect to our hypothesis. We consider an experiment constraint time of 1000 time units and the same experiment setup. 
One can observe in experiment results of Figure~\ref{fig:bee_scalability}, that an individual robot cannot find the optimal location. Similarly, a group of three is very unlikely (probability less than 10\%) to 
succeed.
Increasing the size of the swarm 
improves the probability of gathering into an optimal cluster, and becomes almost certain with more than 27 robots. 
We further note that the waiting time related to the local temperature in cluster, influences scalability -- if waiting time is shorter, probabilities generally increase.

\subsection{Emergency Response in a Smart City}
\label{sec:evalscenario}

We consider a disaster scenario in an urban environment, where communication infrastructure is disabled and parts of the city may be unsafe; search and rescue must be performed for stranded victims within the city.
To this end, autonomous UAVs are dispatched to locate and rescue victims~\cite{fse17}. 
 The city is naturally physically comprised of buildings, roads, squares etc, which can be considered as discrete \emph{locations} where active victims (or UAVs) at any time may reside. 
 UAVs and victims are the two main classes of autonomous agents we consider. 
UAVs move between adjacent locations in the city searching for victims, while it is known beforehand that victims themselves may move between specific locations due to local conditions (e.g., an estimation of where victims may be stranded). 
If victims are found in the same location with a UAV at the same time, they are said to be \emph{saved}, as the UAV successfully detects a victim and alerts rescue personnel of its location. 

As disaster scenarios are highly dynamic and uncertain, individual UAVs performing the search and rescue operation may \emph{crash} due to the disaster-struck volatile environment.
For example, arbitrary obstacles, falling debris or building collapses may introduce hazards for the UAVs operating in the city in a dynamic and unknown way.  
This can be mitigated by communication infrastructure on board the UAVs, which attempts to communicate hazardous local situations encountered by a UAV to neightbouring ones. If communication is successful, neighboring UAVs avoid entering the corresponding location for a defined period of time. 
Besides dynamic obstacles, search and rescue is naturally an energy-consuming task for UAVs,
which have to periodically recharge their batteries at charging stations during the mission. However, battery consumption is not entirely predictable -- strong local wind conditions for instance, lead to UAV's rotors consuming variable power. 
 In our scenario, if a UAV cannot manage to get to a charging station before its battery runs out, it is considered as \emph{out of battery} and is disabled. 


The search and rescue scenario described, implies significant analysis challenges. Firstly, the scenario takes place in a large physical space, where autonomous agents interact with space-bound facilities, such as charging stations.
Secondly, the dynamic environment of a disaster-striken city implies that uncertainty is a key component -- locations of victims may be partially known and behavior of the active agents is governed by probability distributions.
Thirdly, from a system designer perspective, the deployment setup is critical. Choosing where to place the active UAVs (or charging stations) in the large spatial space for instance, will greatly affect their effectiveness. Finally, different coordination or communication mechanisms for UAVs might affect the mission achievement and must be evaluated. 
The overall system goal concerns the entire CPSS that stranded victims, UAVs, charging stations and their interaction protocol induce, and entails rescuing victims. Specifically, we consider a complex goal where:

\begin{displayquote}
``The number of UAVs crashing or running out of battery should be less than one in the case of less than half of the victims being rescued''.
\end{displayquote}

\noindent
System design questions in such an emergency response scenario typically seek to investigate the effect of different deployments given a particular city, initial conditions and system goal. Specifically, design questions in such a scenario typically include:

\begin{itemize}
\item How many UAVs should be deployed to adequately satisfy the system goal?
\item How many charging stations should be deployed to mitigate UAV battery shortages?
\item What is the effect of choosing different initial positions for deploying UAVs?
\item What is the effect of choosing different interaction protocols to mitigate UAV crashes?
\item What is the effect of excluding some parts of the city, thus reducing the search effort for UAVs?
\end{itemize}


\subsubsection{Modeling Design Concerns within the Smart City Space}

\emph{Agents' Spatial Activity}.
Recall that there are two classes of autonomous agents -- victim and UAV. To generate the model of their spatial activity, a topological model of a city is extracted from a CityGML representation, a widely used XML-based standard for the description of city models. While out of scope of the present paper, state-transition structures representing spatial behavior of UAVs and victims are automatically derived and transformed into STA~\cite{fse17,tsigkanos2016architecting}. The process takes as input the movements that UAVs and victims can make within the city, and yields all possible changes in location for each agent given a particular city. Models used are available in an online appendix~\cite{paperstuff}. 
Initial positions (and thus initial states in the respective STAs) of victims can be chosen probabilistically by allowing a distribution over some defined set of initial states. For our disaster scenario, this can be useful if e.g., exact initial positions of victims are not known, but domain knowledge can estimate a part of the city where victims may be located. In contrast, for UAVs, initial positions are given as part of some deployment strategy. 

\emph{Agents' Success and Failure}.
From the system goal, a success state and two failure states are derived. For the victim modeled, 
a ``Saved'' state reflects the success of the \emph{safe} elementary predicate for a single victim entity; if that state is entered by the victim STA, the victim is considered safe. For the UAV modeled, a ``Crashed'' state reflects the UAV's status as \emph{crashed} and a ``OutofBattery'' state reflects the UAV's status as \emph{out of battery}.


\emph{Agents' Interaction and Coordination}.
The mode of communication is assumed to be sourced from communication experts providing a model of the communication protocol as well as its operationalization onboard the UAVs.
For our evaluation purposes and following consultations with UAV experts, we adopt models of \emph{Bluetooth} and IEEE 802.15.4-based \emph{ZigBee}~\cite{zigbee}, 
both commonly used to create networks with low-power radios in industrial scenarios. 
Bluetooth enables low-power short-distance communication, while 
a ZigBee-based setup can provide communication within larger distances of 100 to 500 meters, depending on a power profile and environmental characteristics; communication can reach more distant UAVs through the formation of a mesh network. 
Following consultations with experts, we utilize two simplified interaction STAs of the Bluetooth and ZigBee protocols, which are available in our online complementary material~\cite{paperstuff}.


After obtaining the above STA, an integrated model representing the agent class is exported in XML format as compatible to \emph{$\mathtt{uppaal-smc}$}. 
In this tool, such models are subsequently imported and  the designer specifies the class instantiation based on the scenario at hand -- for our scenario, this entails the number of UAVs and victims as well as their initial positions.
Remaining code implementing function logic is written by the designer and requirements are specified in \emph{$\mathtt{uppaal-smc}$} over the primitives exposed (as per Section~\ref{sec:smc}), and statistical validation is invoked.

\subsubsection{Experiment Setup and Results}


For our experiment setup, victims are initially positioned based on a random distribution in the city. We discuss five scenarios where keeping certain variables fixed, we study the variation in analysis results introduced by another variable: i) the number of UAVs, ii) the number of charging stations, iii) the relative initial positions of UAVs, (iv) the choice of protocols for UAVs as well as (v) excluding certain areas in the city.
For our experiments, we consider a city comprising of 100 Buildings (and according roads, crossroads etc), 
where several UAVs attempt to locate 500 Victims. 
The system goal 
corresponds to the property
$\diamond CrashNum <= 1 \wedge OutofBatteryNum <= 1 \wedge SavedVictimNum >= 250$. 

\begin{figure}[]
  \centering
\includegraphics[width=9cm]{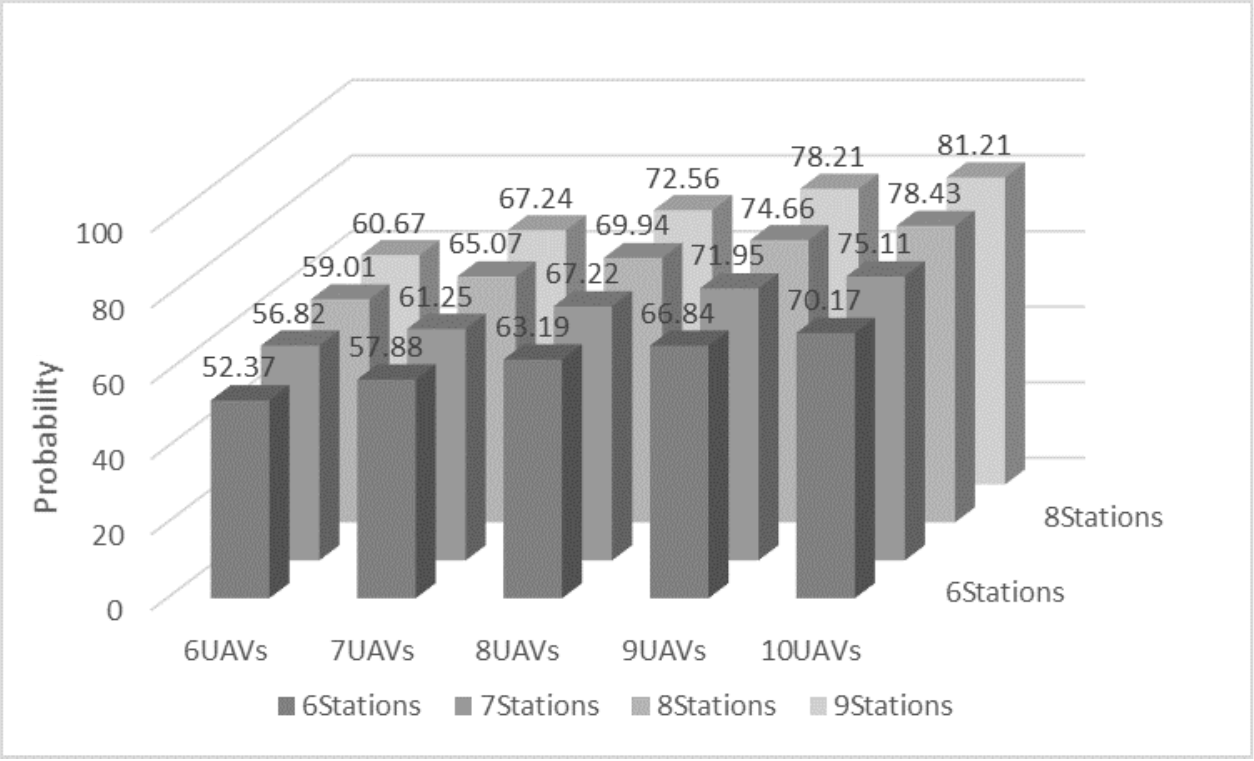}
\caption{Goal satisfaction within the first two design questions.}
\label{fig:exp1}
\end{figure}

\paragraph{Increasing number of UAVs and Charging Stations} For this set of experiments, we stepwisely increase the number of UAVs and charging stations and study their effect on the system goal.
In order to eliminate the influence of UAV initial positions and the choice of communication protocols, we keep UAVs deployed at the same starting point, while communication occurs over Bluetooth.
The typical design question here is that given certain time constraints and a minimum requirement satisfaction desired (i.e., some specified probability threshold), what is the minimal number of UAVs or charging stations that should be deployed in the city to make the probability of satisfying the system goal exceed this threshold. 
Figure~\ref{fig:exp1} shows different configurations of UAVs and charging stations, allowing the designer to choose the optimal; observe how naturally, the probability of successful goal attainment increases by the number of UAVs and charging stations deployed with 95\% confidence. The marginal goal satisfaction gains decrease with additional UAVs, as all are deployed from a single starting point and cannot effectively locate scattered victims.


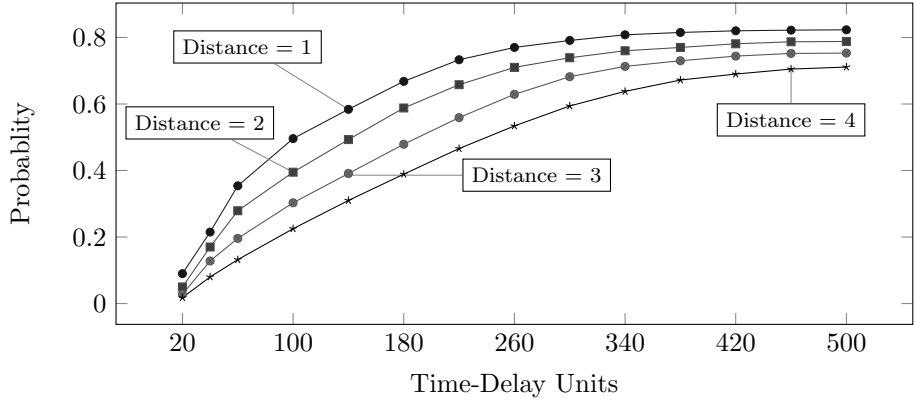
\begin{figure}[h]
  \centering

\tikzstyle{every pin}=[fill=white,
  draw=black,
  font=\footnotesize]
{
\begin{tikzpicture}
\selectcolormodel{gray}
\begin{axis}[
log ticks with fixed point,
  xlabel={Time-Delay Units},
            width=4.75in,
          height=2.3in,
          mark size=1.5,
          xtick={20,100,...,500},
  ylabel={Probablity}
]

\addplot coordinates{
(20,  0.090)
(40,  0.215)
(60,  0.354)
(100,  0.496)
(140,  0.584)
(180,  0.668)
(220,  0.733)
(260,  0.770)
(300,  0.791)
(340,  0.808)
(380,  0.815)
(420,  0.820)
(460,  0.822)
(500,  0.823)
};

\node[coordinate, pin={[pin distance=0.7cm] 120:{Distance = 1}}  ] 
    at (axis cs:140,0.584) {};

\addplot coordinates {
(20,  0.050)
(40,  0.170)
(60,  0.279)
(100,  0.395)
(140,  0.493)
(180,  0.588)
(220,  0.658)
(260,  0.710)
(300,  0.739)
(340,  0.760)
(380,  0.770)
(420,  0.781)
(460,  0.787)
(500,  0.788)
};

\node[coordinate,pin={[pin distance=0.5cm] 120:{Distance = 2}}  ] 
    at (axis cs:100,0.395) {};

\addplot coordinates{
(20,  0.028)
(40,  0.128)
(60,  0.196)
(100,  0.303)
(140,  0.391)
(180,  0.479)
(220,  0.559)
(260,  0.629)
(300,  0.682)
(340,  0.713)
(380,  0.730)
(420,  0.744)
(460,  0.752)
(500,  0.753)
};

\node[coordinate,pin={[pin distance=1.5cm] right:{Distance = 3}}  ] 
    at (axis cs:140,0.386) {};

\addplot coordinates{
(20,  0.018)
(40,  0.080)
(60,  0.132)
(100,  0.225)
(140,  0.310)
(180,  0.389)
(220,  0.466)
(260,  0.534)
(300,  0.594)
(340,  0.638)
(380,  0.672)
(420,  0.690)
(460,  0.705)
(500,  0.711)
};

\node[coordinate,pin=below:{Distance = 4}] 
    at (axis cs:460,0.705) {};

\end{axis}
\end{tikzpicture}
}
\caption{Goal satisfaction over different distances between UAV and victims.
}
\label{exp2}
\end{figure}

\paragraph{Decreasing UAV-Victim distances.} For this set of experiments,
we control the random assignment of initial positions of UAVs in order to be in certain bounds with respect to distances of UAVs-Victims. Distance is denoted by hops between \emph{possible} victim positions and initial UAV positions. We keep the UAV-Victim distance distribution within certain bounds and study its effect on the satisfaction of the system requirement. We deploy randomly 8 charging stations in locations which are kept constant throughout experiments and 8 UAVs utilizing Bluetooth for communication. Actual positions of UAVs and Victims are chosen randomly inside the city, but they adhere to specific UAV-Victim distances. Results are displayed in a probability diagram in Figure~\ref{exp2} with respect to various distance choices. The cumulative probability (lines in Figure~\ref{exp2}) refers to the probability that \emph{time-delay units} is less than or equal to a value on the X-axis (i.e., time units). 
A typical example of this is that the probability can reach around 0.5 representing estimated requirement satisfaction in this configuration with 95\% confidence within a distance of 2. Our result is consistent with the hypothesis based on the requirement -- intuitively, if UAVs are deployed closer to where victims may be located, the requirement is more likely be satisfied and within fewer time units. 

\begin{figure}[]
  \centering
\includegraphics[width=12cm]{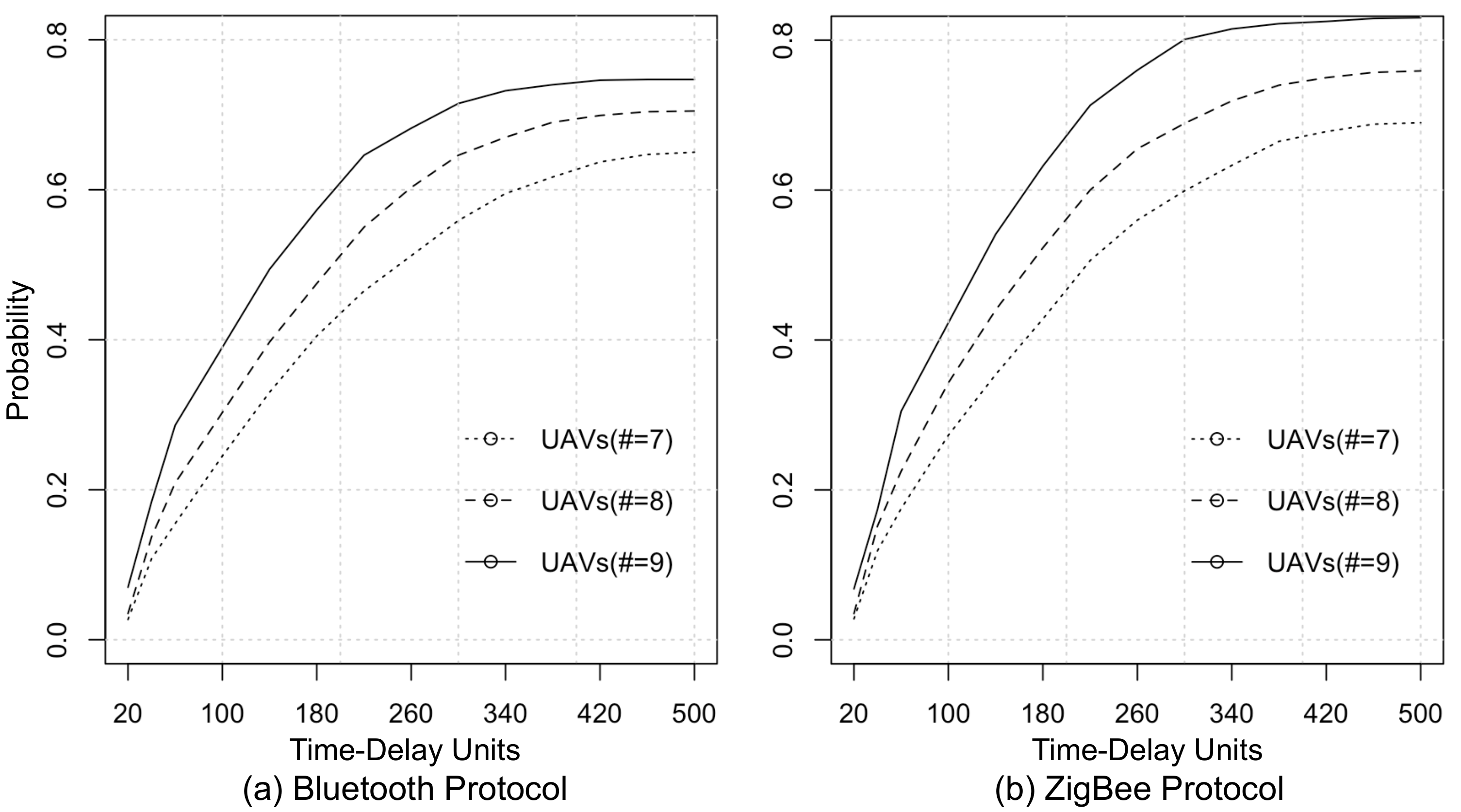}
\caption{Goal satisfaction over different interaction activities.}
\label{fig:interaction}
\end{figure}

\paragraph{Effect of Communication Protocol.}
In this set of experiments, we study the effects of the adoption of Bluetooth and ZigBee for UAV communication. For each experiment, we deploy randomly 6 charging stations and stepwisely increase the number of UAVs from 6 to 8 to compare the degree of satisfaction of goals over the two protocols respectively. Recall that ZigBee is expected to outperform Bluetooth. In Figure~\ref{fig:interaction}, the quantification of this effect is illustrated, and the ZigBee advantage is more evident as the number of UAVs increases. 

\begin{figure}[]
  \centering
\includegraphics[width=8cm]{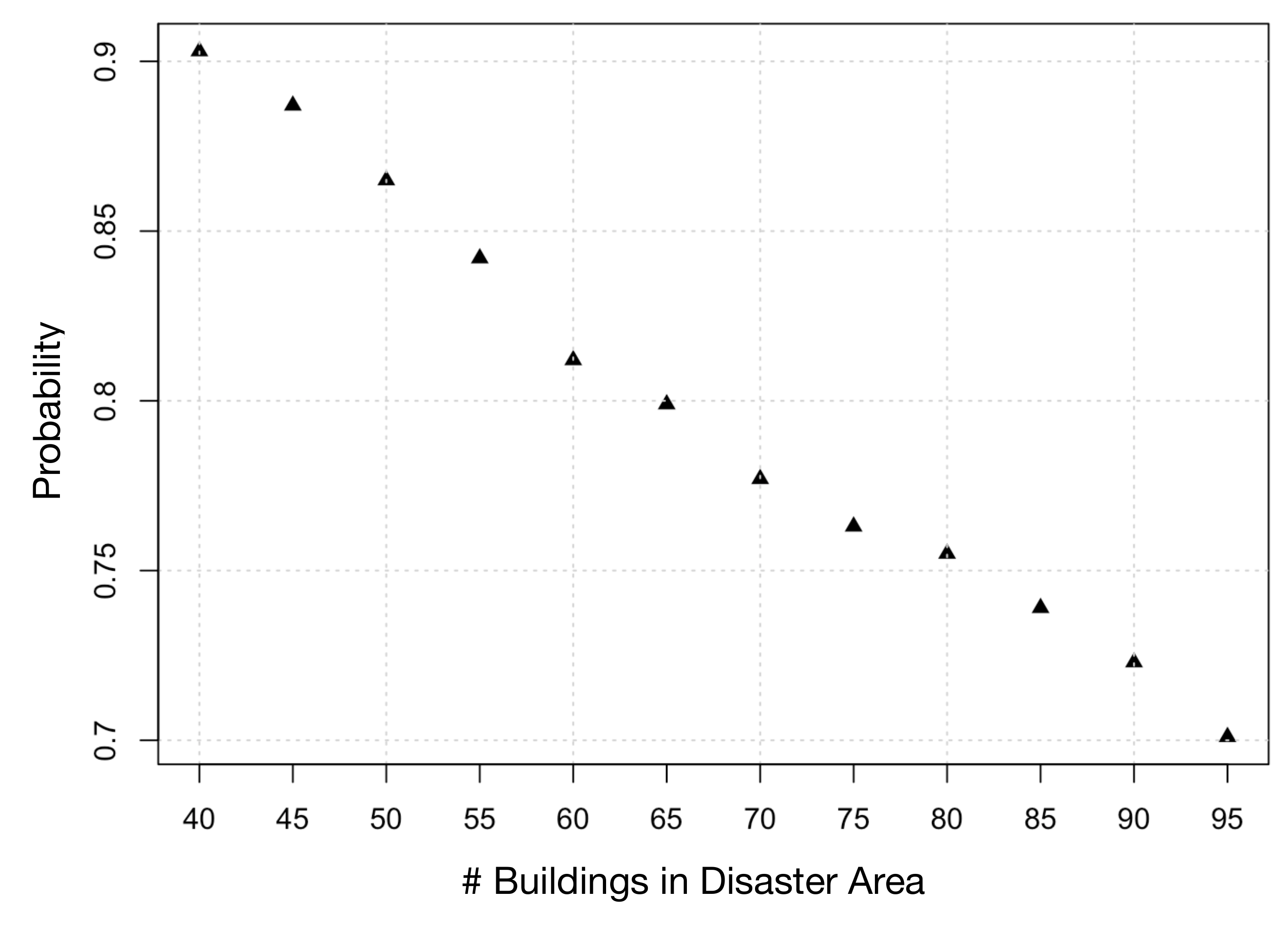}
\caption{Goal satisfaction over different spatial activities.}
\label{fig:spatial}
\end{figure}


\paragraph{Change of the Spatial Concern.}
In this set of experiments, we randomly deploy 8 constant charging stations and 8 UAVs from the same starting point to study the impact of the limiting the size of the spatial activity model. As the number of buildings in the disaster area decreases, indicating zones that the UAVs do not search, 
goal satisfaction is increased (Figure~\ref{fig:spatial}). 
This is consistent with the intuition that UAVs locate victims in a compact area more easily with less  risk of running out of battery and crashing.

\subsection{Discussion}

As evident by the modelling and analysis of the two case studies presented, our approach enables quantitative evaluation of goal satisfaction 
based on design time decisions on deployment choices and individual concern substitution that are crucial in the requirements engineering process.

The analysis workflows inherent in our approach, represent a general solution
schema that can be used to tackle different kinds of CPSS models
whose key design dimensions are spatial activity in the physical space and interaction, while their system mission can be specified in terms of success-failure states.
Straightforward examples of CPSS systems can be smart buildings or cities, with scenarios ranging from disaster management or  infrastructural maintenance to robotic applications. However, depending on
the specific scenario, different extensions may be required, and the general modeling discipline presented would need to be enriched with domain-specific information by the designer (e.g. timing aspects). However, we believe that in the present paper we lay the foundations for analysis of a multitude of CPSS.

Notice that the system development is facilitated since experimentation within the early design process is possible. For example, switching between different spatial models -- while keeping other models stable -- entails solely invoking the composition procedure over the other spatial models.
Switching between different spatial layouts was not demonstrated in the evaluation, since it would be unfeasible to demonstrate quantitative comparisons between models (as each city would be different).
Instead, we studied the impact of the limiting the size of the same spatial activity model, by decreasing the size of the search area in a controlled manner. 

Regarding  performance aspects, results of our evaluation indicate the feasibility of the approach (i.e. with respect to traditional, explicit-state verification).
We additionally report on additional experiments over the disaster scenario case and investigate how analysis times are affected by enlarging scope of spatial activity and numbers of behaviour models considered. Consider a randomly generated city with 500 buildings, where 1000 victims are located and 20 UAVs are deployed to search for them. The analysis time required for this scenario is nearly 33 minutes. Although the analysis time is relatively large, analysis itself is a design time activity thus such performance is not an issue. Overall, we believe that the analysis results signify that our approach is scalable even to larger models and indeed fit for design time requirements validation of CPSS.
The advantage of statistical model checking is evident; analyses for the experiments considered would be infeasible with traditional, explicit-state model checking verification. This is because both the number of active agents (UAVs and victims) as well as the city sizes are large.

\section{Conclusions}
\label{sec:conclusions}

Within the context of complex cyber-physical space systems, support for early requirements validation is crucial to the design process. 
To this end, we outlined a systematic approach to high-level reasoning through separation of key recurrent system concerns and formally defined integration of models that capture them. 
Our contribution consists of a framework unifying existing techniques for the engineering of cyber-physical space systems through semi-automation of model integration,
enabling design-time validation through statistical model checking.
The proposed approach has been applied to two case studies that confirm that design can be generally decomposed through modeling concerns corresponding to spatial and interaction activities and requirements. 
While we defer a thorough discussion of assumptions and limitations to future work, 
we plan to investigate system requirements expressibility and influence on system requirements from other concerns, such as changes on geographical or spatial layout, or planning. 
Additionally, we plan to consider cyber-physical domain-related particulars like sensing and actuation.


\section{Acknowledgment}

Research partially supported by the National Natural
Science Foundation of China under Grant Nos. 61620106007 and 61751210, as well as Lise Meitner FWF Austria project M 2778-N ``EDENSPACE''.


\bibliographystyle{IEEEtran}
\bibliography{paper}

\end{document}